# SCIENTIFIC REPORTS

natureresearch

**OPEN** 

# Three cooperative mechanisms required for recovery after brain damage

D. Berger[1], E. Varriale[2], L. Michiels van Kessenich[1], H. J. Herrmann[3] & L. de Arcangelis[4*]

Stroke is one of the main causes of human disabilities. Experimental observations indicate that several mechanisms are activated during the recovery of functional activity after a stroke. Here we unveil how the brain recovers by explaining the role played by three mechanisms: Plastic adaptation, hyperexcitability and synaptogenesis. We consider two different damages in a neural network: A diffuse damage that simply causes the reduction of the effective system size and a localized damage, a stroke, that strongly alters the spontaneous activity of the system. Recovery mechanisms observed experimentally are implemented both separately and in a combined way. Interestingly, each mechanism contributes to the recovery to a limited extent. Only the combined application of all three together is able to recover the spontaneous activity of the undamaged system. This explains why the brain triggers independent mechanisms, whose cooperation is the fundamental ingredient for the system's recovery.

Brain strokes, caused by an interruption in blood supply[1], are a leading cause of adult disability. The reduced oxygen supply in localized regions, perturbing energy-dependent metabolisms, ultimately results in neuronal death[2], also affecting the activity over large regions. Experimental observations[3] evidence that the ischemic core, i.e. the region where cells are irreversibly injured, is surrounded by a region experiencing reduced blood flow, which is termed *penumbra*[1]. This is the prime region where most of the compensatory recovery mechanisms take place[4]: (1) Structural recovery after a stroke leads to the reactivation of activity dependent plasticity[5]. (2) Neurons in the penumbra exhibit an increased firing rate[6], named hyperexcitability, because of a reduced efficacy of $GABA_A$ synapses[7]. (3) Growth-associated genes important in axonal sprouting are expressed shortly after a stroke[8,9] leading to long distance axonal sprouting near the ischemic injury[10,11] Synaptogenesis is observed to depend on neuronal activity[4]. Although a rich description of different physiological mechanisms triggered after a stroke and during the recovery has been provided experimentally, how the observed mechanisms each work towards the recovery is still unknown. Computational models of neural networks, which have been designed to understand the functioning of the brain, have up to now not looked into the implications of brain injuries and the recovery mechanisms[12]. Understanding the role of each mechanism in the recovery is of fundamental importance to understand how the brain reacts to such injuries and opens the way to boost the recovery. Here we address this open problem by considering the main experimental observations in a neuronal network. We unveil the physiological role played by each mechanism and evidence that only the interplay among them leads to full recovery.

Unlike higher brain functions, the spatio-temporal features of spontaneous activity in healthy brains, are not only well described by experiments[13–21] but also confirmed by neuronal models[22–25]. In particular, spontaneous activity exhibits a characteristic spatio-temporal pattern, critical avalanches, first observed in organotypic cultures from coronal slices of rat cortex[13]. The distribution of avalanche sizes consistently exhibits the scaling $P(s) \propto s^{-1.5}$. Critical avalanche dynamics is also found in cortical activity of awake moneys[18] as well as in human fMRI[26] and MEG recordings[20]. Inspired in self-organized criticality (SOC)[27,28], a model implementing the main physiological features of real neurons[22,29] was able to reproduce the experimental MEG spectra and avalanche statistics. Neuronal network models implementing both short- and long-term synaptic plasticity also reproduce neuronal avalanche behaviour at a tunable parameter critical value[23,25]. Spontaneous activity plays a relevant role in the response of a brain to a given task. Arieli et al.[30] have evidenced, by combining optical imaging with electrophysiological techniques, that in awake animals the response to repeated presentations of the same stimulus depends on the instantaneous cortical state of ongoing activity. As a consequence, the observed disabilities in brain functions after a stroke may be intimately related to alterations in the resting state. Here we address the question how a damage affects the spontaneous activity of a neuronal network and how the system is able to recover. In order to identify more clearly the consequence of a stroke, we investigate also the case of a damage not localized but affecting the same number of neurons randomly distributed throughout the system. This last damage condition is reminiscent of brains affected by advanced atherosclerosis or senile dementia. The physiological mechanisms observed to take place after a stroke are implemented and their role in re-establishing the healthy critical state is discussed evidencing the key role of cooperative effects.

[1]Computational Physics for Engineering Materials, IfB, ETH Zürich, CH, Zürich, Switzerland. [2]Physics Department, University of Naples Federico II, 80125, Naples, Italy. [3]PMMH, ESPCI, 7 quai St. Bernard, 75005 Paris, France and Departamento de Fisica, Universidade Federal do Ceará, 60451-970, Fortaleza, Ceará, Brazil. [4]Dept. of Engineering, University of Campania "Luigi Vanvitelli", 81031 Aversa (CE), INFN sez. Naples, Gr. Coll., Salerno, Italy. *email: lucilla.dearcangelis@unicampania.it

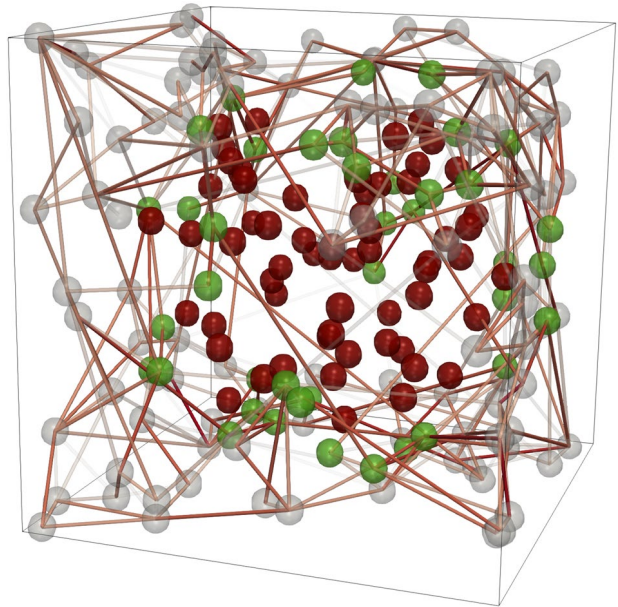

**Figure 1.** Configuration with localized damage. Configuration of a neural network ($N = 200$) where a 30% localized damage is implemented: The damaged neurons (red) are surrounded by neurons of the penumbra (green). Unaffected neurons are grey. The connections of all damaged neurons are removed. The penumbra size is 20%$N$.

## Results

A configuration of the neuronal system with $N$ neurons is generated with a scale free network of connections. Synaptic strengths, initially assigned with a random value, are sculpted by activity dependent plasticity (see Methods). All systems exhibits critical avalanche dynamics (see Supplementary Information). Structural damage is introduced into the experienced system, i.e. at the end of plastic adaptation. The network does not represent the entire brain, but only the functional region wherein the lesion occurs. Indeed, the modular brain structure[31] has a relevant role in spontaneous activity and brain damage is often not constrained into a single functional area, as it is the case of atherosclerosis and senile dementia. In order to shed light on the role of each mechanism in the recovery process, here we focus on the particular case of an intra-module damage. The stroke can affect a variable fraction of neurons, up to 50% of the region, and has a spherical shape (Fig. 1)[32]. Similar results are obtained for ellipsoidal damage and for a different neuronal model[22]. All neurons in the damaged area are removed together with all incoming and outgoing synapses. In addition to the localized damage, we also analyse the case of diffuse damage (see Supplementary Information), where we remove the same number of neurons, with all incoming and outgoing connections, but choosing the neurons at random in the system.

**Activity after damage.** In order to monitor the effects of damage, we analyse the statistics of spontaneous activity, namely the avalanche size distribution. The avalanche size is measured either as the number of firing neurons, or as the sum of potential variations induced at post-synaptic neurons. In healthy systems the largest avalanche involves all neurons $N$. For both damages, the largest avalanche is affected by the irreversible death of neurons and therefore can involve only all surviving neurons. In order to compare the avalanche statistics for damages of different magnitudes, we evaluate the rescaled avalanche size distribution $P(s)s^{1.5}$ as function of $s/c$, the size rescaled by the cutoff, defined as the value of avalanche size for which only 0.1% of avalanches are larger. In the following the avalanche size is measured in terms of potential variations.

Interestingly, in the case of diffuse damage (Fig. 2b) the scaling of the distribution is weakly affected in the middle size range showing a small increase of intermediate size avalanches. Conversely, for the localized damage case (Fig. 2a) the distributions are sensibly modified exhibiting an excess of intermediate and large size avalanches, which indicates that the system behavior tends to become supercritical. In order to further characterize the consequences of damage we measure the average avalanche size versus $c$ for each configuration. In the undamaged case, where the distribution follows a power law scaling $s^{-1.5}$ up to the cutoff fixed by the system size, the average avalanche varies with the cutoff as $<s> \propto \sqrt{c}$. The scatter plot in Fig. 2c (inset) shows that this scaling is indeed verified for different sizes of diffuse damage, namely for increasing damage the average size is solely tuned by the cutoff, confirming that diffuse damage slightly affects the critical behaviour. Conversely for increasing localized damage the average size decreases more slowly, as $c^{0.2}$, which implies the emergence of an excess of large avalanches compensating the reduced cutoff. In order to further understand the origin of the strong increase in large avalanches observed for localized damage, we monitor the average firing rate for a given rescaled avalanche size, $s/c$ (Fig. 2c). This is obtained by evaluating the average firing rate for all neurons active during an avalanche of size $s$. In agreement with previous results, the average firing rate does not appreciably change for diffuse damage, whereas strongly increases in the range of intermediate sizes after a stroke. This counter-intuitive result can be understood by considering that, because of the death of neurons and all their ingoing and outgoing synapses, activity involves only the surviving neurons which fire more often. In conclusion, diffuse damage has a limited effect on spontaneous activity, basically rescaling it to a smaller system size. Conversely, the localized damage has a much stronger impact on the system which goes well beyond rescaling activity because of the reduced number of neurons. Surviving neurons exhibit a higher firing rate causing an unexpected excess of large avalanches leading to a supercritical scaling behavior for the distribution.

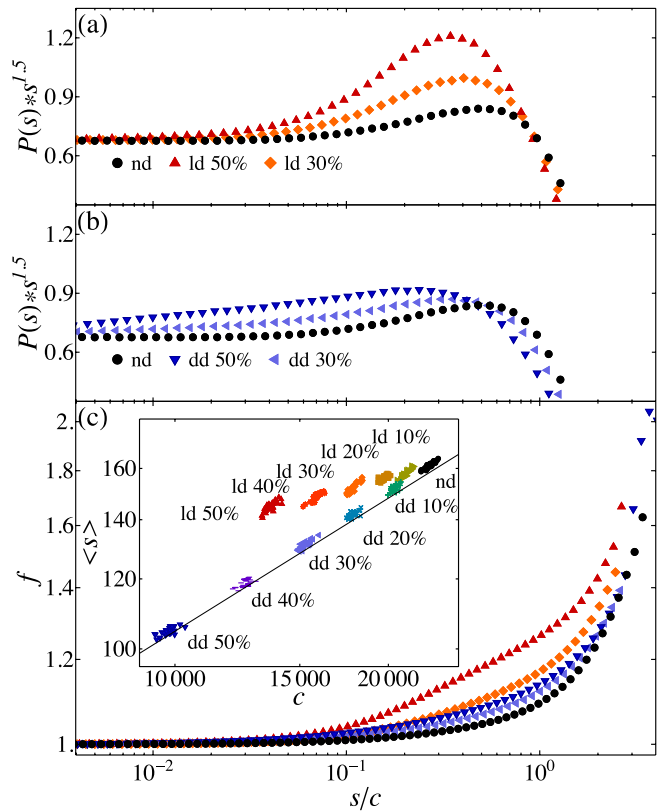

**Figure 2.** Activity after damage. Rescaled avalanche size distribution $P(s)s^{1.5}$ versus $s/c$ for different localized (**a**) and diffuse damages (**b**). In both panels the distribution for undamaged networks is also shown. (**c**) Average firing rate for different avalanche size $s$ plotted versus the rescaled size $s/c$. Inset: Scatter plot of the average avalanche size versus the cutoff $c$ for different localized and diffuse damages and for the undamaged case (black). Each symbol represents a different network configuration. The straight line indicates the relation $s \propto \sqrt{c}$.

**Recovery after damage.** In order to implement the three different recovery mechanisms, we define the penumbra as 20% the number of neurons $N$, placed in the shell surrounding the lesion. We test the recovery performance of the system under the main three mechanisms detected experimentally (Fig. 3). In particular, (1) synaptic plasticity is reactivated following the same rules as in the training period, with a maximum number of plastic adaptations $N_{rec} = 20000$, and affects only synapses whose pre-synaptic neuron is the penumbra. (2) Hyperexcitability involves neurons belonging to the penumbra and implies a change in the firing rate of both excitatory and inhibitory neurons. This is implemented by decreasing (increasing) the firing threshold of excitatory (inhibitory) neurons by the same $\delta t$. (3) Synaptogenesis is implemented at the end of each avalanche by adding a new connection to all presynaptic neurons belonging to the penumbra which fired more than once during the avalanche. The new synapse connects to a post-synaptic neuron not belonging to the penumbra and at an average distance larger than the initial connections[10,11]. The recovery performance is analysed for different values of the average distance and different numbers of sprouted synapses, expressed in terms of the percentage of the number of connections killed by the stroke.

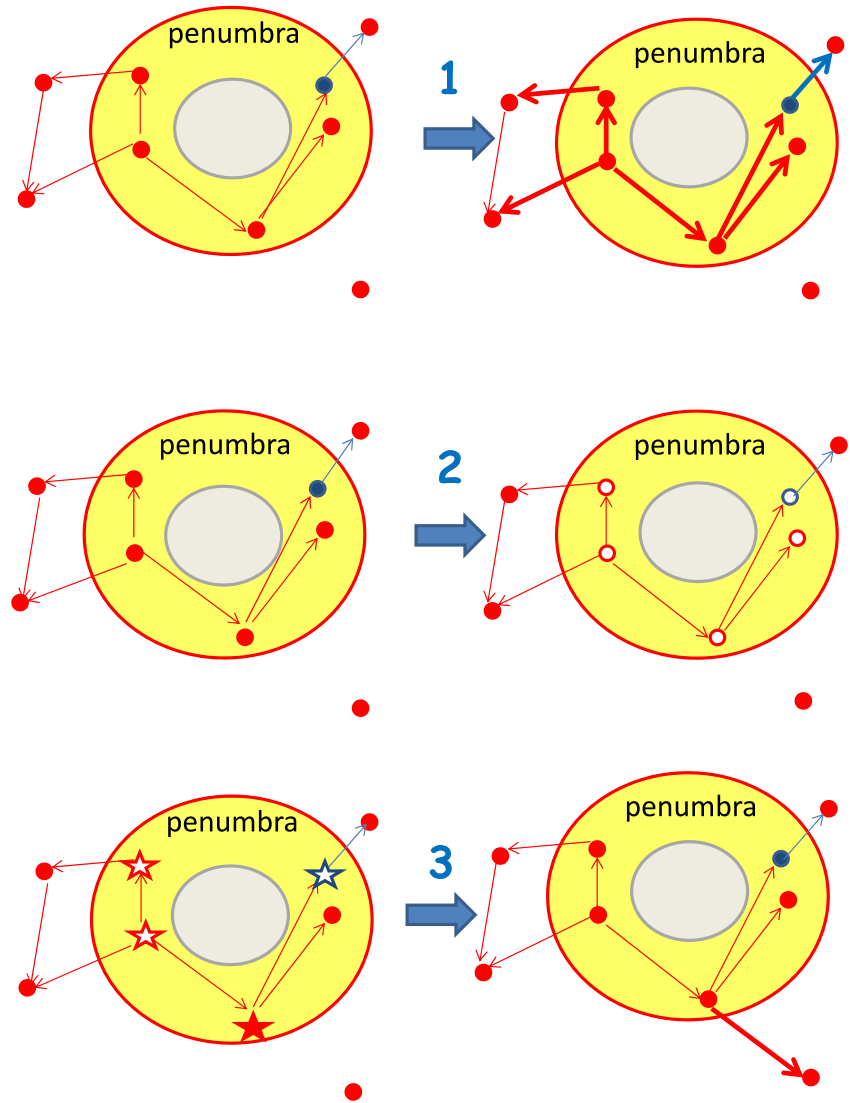

**Figure 3.** Illustration of the three recovery mechanisms. Excitatory neurons and synapses are shown in red, inhibitory ones in blue. The penumbra (yellow area) is a circular shell around the stroke (grey). (1) Plasticity reactivation: Outgoing synapses from neurons belonging to the penumbra are strengthened. (2) Hyperexcitability: Neurons in the penumbra have their firing threshold modified (empty symbols). (3) Synaptogenesis: The neuron in the penumbra that fired at least twice in the previous avalanche (full star) sprouts a new connection to a neuron outside the penumbra.

We first analyse the effect of reactivating the plastic adaptation of synapses outgoing from neurons belonging to the penumbra. Figure 4 shows that plasticity has a little effect in the recovery of the undamaged system for all three criteria. This limited role in the recovery is observed also increasing the temporal duration of the plastic adaptation $N_{rec}$ to 40000 stimulations and the spatial extension of the penumbra to 40% the number of neurons (see Supplementary Information). It therefore appears that plasticity alone is unable to act on any monitored quantity to retrieve the critical behavior of avalanche activity. Next, we implement hyperexcitability by varying the firing thresholds of excitatory and inhibitory networks in the penumbra by $\delta t$. Small variations of the threshold indeed leave the average firing rate totally unaffected but reduce the excess of large avalanches, as well as the value of the cutoff (inset of Fig. 4). This counter-intuitive effect can be understood by considering that excitatory neurons, by firing more often, weaken the signal to connected neurons. Interestingly, for values of $\delta t$ larger than 0.2 the scaling behaviour of the distribution is strongly affected and a further increase in the firing rate is observed bringing the system away from criticality. Larger penumbra sizes do indeed help the recovery, however penumbras larger than 20% are not realistic experimentally (see Supplementary Information).

Finally, the sprouting of new connections, by increasing the connectivity, has the strongest effect on the recovery of critical scaling properties: The excess of large avalanches and the average firing rate are reduced but also a slight increase in the cutoff is observed, The effect improves with long-ranged connections for all three criteria, since larger areas of the system are involved in the recovery. The number of newly sprouted synapses is also a relevant parameter, providing the optimal performance already for the creation of about 5% the number of synapses killed by the stroke (see Supplementary Information). None of the three mechanisms at work alone after a stroke is able to fully recover the behavior of the undamaged system. In Fig. 4 we compare the three mechanisms: Synaptogenesis has the stronger effect in the recovery of the scaling of the size distribution, the firing rate and the cutoff. However, hyperexcitability is able to modify the dependence of the average size on the cutoff. Interestingly, only the application of the three mechanisms in a combined parallel way is able to recover the critical behaviour of avalanche activity. Because of irreversible cell death, the cutoff in avalanche size extension is always limited. However, deviations observed from the healthy state are drastically reduced for the combined three criteria.

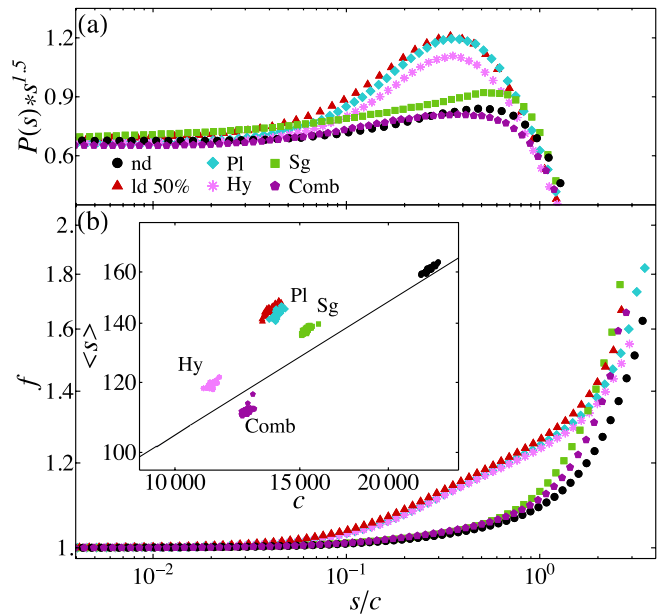

**Figure 4.** Activity after recovery. (**a**) Rescaled avalanche size distribution as function of the rescaled size $s/c$ for the damaged (50%) and undamaged networks, for the three recovery mechanisms applied separately and in a combined way (Comb): Plasticity (Pl), hyperexcitability (Hy) and synaptogenesis (Sg). The parameters are: $N_{rec} = 20000$, $\delta t = 0.2$, $r_0 = 9$. (**b**) Firing rate vs. $s/c$ for undamaged, localized damage (50%) and the three recovery mechanisms as in (**a**). The inset shows the scatter plot of the average avalanche size vs. the cutoff for the same color code.

The existence of cooperative effects is confirmed by the implementation of the three mechanisms in a random sequential order (see Fig. 5). Indeed similar recovery performance is observed if the three mechanisms are not applied simultaneously but one after the other on the same system. Conversely, the recovery observed for an ensemble of networks where one of the three mechanisms is applied at random and independently of the others, is drastically reduced. This result confirms that is the interaction among the three mechanisms the main ingredient for recovery. These three mechanisms are observed to take place after a stroke but for completeness we verified their efficacy also after diffuse damage (see Supplementary Information). In this case the penumbra is identified as a percentage of surviving neurons. Interestingly, we find that only plasticity and synaptogenesis would partly help the recovery, whereas hyperexcitability has a counterproductive effect. The three mechanisms combined bring the system even farther from the healthy state, confirming that these mechanisms are problem specific and triggered by the brain only after a stroke.

## Discussion

A stroke has important consequences on the activity of a neuronal system, which go beyond the functional disabilities but affect also the resting state. Indeed the localized death of a number of neurons has a strong impact on the spontaneous activity of the system and, more precisely, on the average size of avalanches or bursts, their size distribution and their average firing rate. Interestingly, it is the localization of the damage that is responsible for such deviations, not observed in the case of a diffuse damage of the same size. Experimental observations indicate that three main mechanisms take place after a stroke, here we unveil their role. Synaptogenesis performs the most important task since it is able, by increasing the connectivity, to distribute the signal over a wider region and therefore reduce the firing rate of surviving neurons. As a consequence, also the excess of large avalanches and their average size are drastically reduced. At the same time, hyperexcitability, by facilitating the neuronal firing in the penumbra, acts on avalanches of all sizes reducing, both, the excess of large avalanches and their average size. By itself, hyperexcitability is unable to reduce the average firing rate and its presence could appear counterproductive. However, hyperexcitability is intimately related to synaptogenesis since, increasing the firing rate of excitatory neurons in the penumbra, it facilitates the sprouting of new synapses partially restoring the connectivity. Plasticity by itself appears to be unable to recover healthy behavior but is important in sculpting the strength of new synapses. We stress that both synaptogenesis and hyperexcitability are local effects, involving mostly neurons in the penumbra. Conversely, plasticity can have long-range effects involving in a cascade also connections far from the penumbra or even beyond the functional region. This mechanism could then be effective at the scale of the entire brain, leading to even contralateral effects balanced by ipsilateral compensation. It is only the combined application of all three mechanisms, either sequentially or in parallel, that leads to recovery towards healthy conditions. These results suggest that experiments on animals after an induced stroke by the combined use of receptor antagonists, as bicuculline or picrotoxin, which reduce and enhance the activity of inhibitory and excitatory neurons, respectively, may confirm that controlling the excitability level in the system may enhance synaptogenesis and help the recovery.

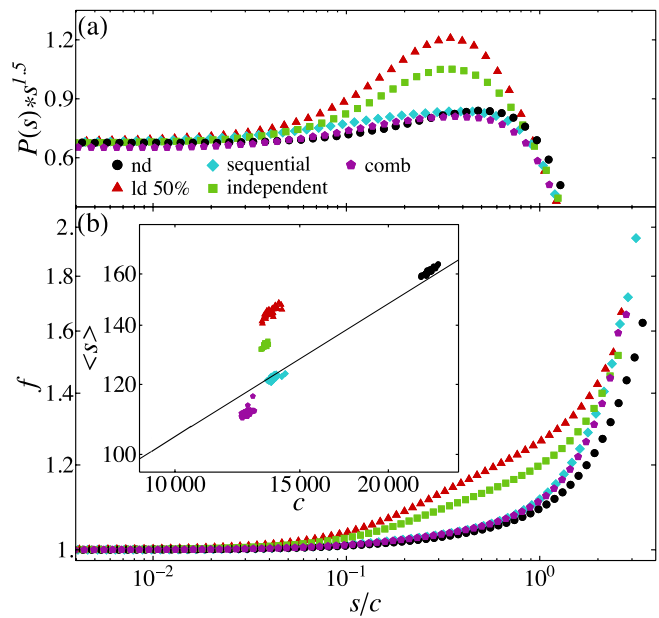

**Figure 5.** Comparison of recovery mechanism implementation. (**a**) Rescaled avalanche size distribution as function of the rescaled size *s/c* for the damaged (50%) and undamaged networks, for the three recovery mechanisms applied independently of each other, sequentially and in a parallel way. The parameters are: $N_{rec} = 20000$, $\delta t = 0.2$, $r_0 = 9$. (**b**) Firing rate vs. *s/c* for undamaged, localized damage (50%) and the three recovery mechanisms as in (**a**). The inset shows the scatter plot of the average avalanche size vs. the cutoff for the same color code.

## Methods

**Network model.** We consider a network with *N* neurons, placed randomly in a three-dimensional cube. Synaptic connections are directed and a fraction $p_{inh} = 0.2$ of neurons inhibitory. The network of out-going connections is scale-free, sampled according to the experimental distribution of functional networks[33], $n(k_{out}) \propto k_{out}^{-2}$ with $k_{out} \in [k_{min}, k_{max}] = [5,100]$. Since inhibitory neurons are hubs[34], they are assigned among neurons with large out-degree. The probability to establish a connection depends on the spatial distance *r* between two neurons according to $p(r) = \frac{1}{r_0} e^{-r/r_0}$[35], with $r_0 = 3$. Initial synaptic strengths $\omega_{ij}$ are chosen to be uniformly distributed within the interval [0.04, 0.06]. Every neuron *i* is characterized by a membrane potential $v_i$, whose initial values are uniformly distributed in the interval [0.1, 0.4]. At each time step all neurons with a potential exceeding a threshold value $v_i \geq v_c = 1.0$ fire, causing a change in the potential of the connected post-synaptic neurons *j* as[25]

$$v_j(t+1) = v_j(t) \pm v_i(t) u_i \omega_{ij}$$
$$u_i(t+1) = u_i(t)(1 - \delta u)$$
$$v_i(t+1) = 0 \qquad (1)$$

where $u_i$ stands for the amount of releasable neurotransmitter, which initially is set equal to one for all neurons and for simplicity is the same for all synapses of neuron *i*. The amount of releasable neurotransmitter reduces by roughly 5% after a spike[36,37], therefore $\delta u = 0.05$. The plus or minus sign in Eq. (1) stands for excitatory or inhibitory synapses, respectively. After a neuron fires, it remains in a refractory state for one time step during which it is unable to receive or send any signal. The important feature generating the critical dynamics of the resting state is the slow recovery process of $u_i$ combined with the the fast depletion during activity[23,25]. The recovery of synapses takes a time of the order of seconds[38] whereas synaptic transmission is of the order of milliseconds. Also stimulation by spontaneous vesicle release occurs at a rate of hundreds of milliseconds to seconds[39]. This consideration allows us to assume that during an avalanche neurotransmitter resources are not retrieved. After an avalanche ends, $u_i$ of each neuron *i* is increased by an amount $\Delta u_r$, which determines if the network will be in a subcritical, critical or supercritical state. The value $\Delta u_r = 3 \cdot 10^{-5}$ sets the system for $N = 40000$ neurons at criticality, i.e. the distribution behaves as $s^{-1.5}$ and the cutoff scales as the system size. Moreover, activity is kept ongoing by adding a small $\delta v = 0.1$ to a random neuron (see Suppl. Figure 2).

The synaptic network structure is shaped by activity-dependent plasticity: Whenever a neuron $i$ fires, all weights of outgoing synapses, connecting $i$ to an active post-synaptic neuron $j$ are strengthened according to the voltage variation induced in the post-synaptic neuron

$$\omega_{ij}(t+1) = \omega_{ij}(t) + \delta\omega_{ij} \tag{2}$$

where $\delta\omega_{ij} = \alpha(v_j(t+1) - v_j(t))$ and $\alpha = 0.0003$ only determines the rate of this plastic change. At the end of each avalanche unused synapses are weakened by the average increase per bond $\omega_{ij} = \omega_{ij} - \sum \delta\omega_{ij}/N_s$ where $N_s$ is the total number of synapses. Synaptic plasticity efficiently weakens loops in the network structure, depending strongly on the length of the refractory period[40,41]. When an avalanche ends, all strengths $\omega_{ij}$ are decreased by the average increase in synaptic strength per bond. Plastic adaptation is implemented for a given number of external stimulations, e.g. $N_p = 10000$. If, however, a synaptic strength reaches zero, the adaptation is interrupted and the synapse is pruned. Data are averaged over 50 configurations of networks with $N = 40000$ neurons.

### Acknowledgements

HJH would like to thank the Agencies CAPES and FUNCAP for financial support. LdA would like to acknowledge the financial support from MIUR-PRIN2017WZFTZP.

### Author contributions

L.d.A. and H.J.H. conceived the research, D.B., E.V. and L.M.v.K. ran the simulations and analysed the results. All authors reviewed the manuscript.

### Competing interests

The authors declare no competing interests.

### Additional information

**Supplementary information** is available for this paper at https://doi.org/10.1038/s41598-019-50946-y.

**Correspondence** and requests for materials should be addressed to L.d.A.

**Reprints and permissions information** is available at www.nature.com/reprints.

**Publisher's note** Springer Nature remains neutral with regard to jurisdictional claims in published maps and institutional affiliations.